\documentclass[aps,prb,twocolumn,amsmath,10pt]{revtex4-1} 
\pdfoutput=1
\usepackage{feynmp, tensor}

\usepackage{ulem}

\usepackage{bbm}

\usepackage{color,graphicx}
\usepackage{bm}

\usepackage{dsfont}
\usepackage{amsmath}
\usepackage{amsfonts}
\usepackage{amssymb}

\usepackage{subfigure}
\usepackage[latin1]{inputenc}

\renewcommand{\-}{\,-\,}

\newcommand{\be}{\begin{equation}}
\newcommand{\ee}{\end{equation}}

\newcommand{\bea}{\begin{equation}\begin{aligned}}
\newcommand{\eea}{\end{aligned}\end{equation}}
\let\oldmarginpar\marginpar
\renewcommand\marginpar[1]{\-\oldmarginpar[\raggedleft\tiny #1]%
{\raggedright\tiny #1}}

\begin{document}

\title{Chaos in a classical limit of the Sachdev-Ye-Kitaev model}
\author{Thomas Scaffidi}
\affiliation{Department of Physics, University of California, Berkeley, CA 94720, USA}
\affiliation{Department of Physics, University of Toronto, Toronto, Ontario, M5S 1A7, Canada}
\author{Ehud Altman}
\affiliation{Department of Physics, University of California, Berkeley, CA 94720, USA}

\date{\today}
\pacs{
}

\begin{abstract}
We study chaos in a classical limit of the Sachdev-Ye-Kitaev (SYK) model obtained in a suitably defined large-$S$ limit.
The low-temperature Lyapunov exponent is found to depend linearly on temperature, with a slope that is parametrically different than in the quantum case: it is proportional to $N/S$.
The classical dynamics can be understood as the rotation of an $N$-dimensional body with a random inertia tensor, corresponding to the random couplings of the SYK Hamiltonian.
This allows us to find an extensive number of fixed points, corresponding to the body's principal axes of rotation.
The thermodynamics is mapped to the $p$-spin model with $p=2$, which exhibits a spin glass phase at low temperature whose presence does not preclude the existence of chaos.

\end{abstract}
\maketitle

\section{Introduction}
The last years have seen a surge of interest in the thermalization dynamics of closed quantum systems. One of the most intriguing outcomes of these activities  has been a deeper and increasingly quantitative understanding of quantum chaos. A striking result, was the proof by Maldacena {\it et. al.} of an upper bound  on the rate at which chaos can develop in a quantum system, as characterized by a Lyapunov exponent $\lambda\le 2\pi  T/\hbar$ ~\cite{Maldacena2016}. Soon after, Kitaev introduced a solvable model of interacting fermions, known as the Sachdev-Ye-Kitaev (SYK) model, which at low temperatures saturates the quantum bound on chaos \cite{PhysRevLett.70.3339,PhysRevB.59.5341,PhysRevB.63.134406,KitaevKITP,Polchinski2016,PhysRevD.94.106002,BAGRETS2016191,BAGRETS2017727} (see also \cite{0305-4470-36-12-340} for a review of earlier work). 

The recent results depart from a long history of studies of quantum chaos in a number of ways. An important new element is the focus on many-body systems rather than on the dynamics of single or few particles. Indeed a major goal of recent studies has been to relate the chaotic dynamics to the transport coefficients which govern the late time hydrodynamic behavior of thermalizing many-body systems\,\cite{Hartnoll:2015aa,Blake2016,Gu2017,Blake2017,Lucas2017,Werman2017}. Such a relation could have important implications for the puzzle of ``strange metals'' \cite{PhysRevLett.69.2975,PhysRevLett.74.3253,2017arXiv170500117S}.

 Another departure from previous work on quantum chaos concerns the relation to a classical limit. Studies of quantum chaos have long highlighted the close link to chaotic classical systems in the limit $\hbar\to 0$ \cite{Berry1978,gutzwiller1991chaos,PhysRevE.50.888}. 
Quantum chaos was in fact defined as the behavior of a quantum system whose classical limit is chaotic.
It was later shown that near the classical limit the assumptions of random matrix theory (RMT), pertinent to quantum chaos, are indeed consistent with classical chaotic dynamics \cite{PhysRevLett.76.3947}. The new results on the chaos bound and the SYK model, on the other hand, are not rooted in the behavior of a limiting classical system. In particular the Lyapunov exponent
$\lambda_\text{bound}=2\pi T/\hbar$ characterizing fast scramblers seems not to have a finite classical limit. 
Nonetheless it is worth asking whether systems exhibiting (maximal) quantum many-body chaos already have unusual features at the classical level.

In this article, we study chaos in a classical limit of the Sachdev-Ye-Kitaev model obtained in a large-$S$ limit, where $S$ refers to the spin-$S$ spinor representation of $SO(N)$.
We calculate numerically the Lyapunov exponent at all temperatures by studying the exponential divergence of a suitably-defined sensitivity, which is the classical limit of an out-of-time-order commutator (OTOC) \cite{larkin1969quasiclassical,COTLER2018318}.
We find that the Lyapunov exponent is linear in $T$ at low temperature, like in the quantum case.
However, the corresponding slope is parametrically different in the classical case than in the quantum case, which raises interesting questions regarding the quantum to classical crossover.

Other interesting properties of the classical SYK model are then studied.
The thermodynamics is mapped to the $p=2$-spin spherical model \cite{PhysRevLett.36.1217}.
The dynamics is identified as the rotation of an $N$-dimensional body around its center of mass, with a random inertia tensor set by the random couplings of the SYK Hamiltonian.
This leads to the identification of $N(N-1)$ fixed points, corresponding to the principal axes of rotation, which span the whole range of energy densities.
A linear stability analysis around these fixed points provides us with a convenient way of calculating a spectrum of local Lyapunov exponents \cite{Abarbanel1991}.

\section{Model}
We start by rewriting the SYK Hamiltonian as a $\mathfrak{so}(N)$ rotor Hamiltonian:
\bea
\hat{H} =  \frac12 \sum_{i<j,k<l} J_{ijkl} \gamma_i \gamma_j \gamma_k \gamma_l = \frac12 \sum_{a,b} \mathcal{J}_{ab} \hat{L}_a \hat{L}_b 
\label{SYKHam}
\eea
Here the $\gamma_i$ ($i=1,\dots,N$) are Majorana fermion operators (with anticommutation relations given by $\{\gamma_i, \gamma_j\}=\delta_{ij}$) and  $J_{ijkl}$ is a completely antisymmetric tensor, whose components are picked randomly with a Gaussian distribution with zero mean and with an energy scale set by $\overline{J_{ijkl}^2} \propto J_0^2$.
We have  defined generalized angular momentum components composed of fermion bilinears $\hat{L}_{ij} =  i \hbar \gamma_i \gamma_j/2$. These operators obey the $\mathfrak{so}(N)$ algebra, $[\hat{L}_{a}, \hat{L}_{b}] = -i \hbar f_{abc} \hat{L}_c$, with $f_{abc}$ the $\mathfrak{so}(N)$ structure constants and where we used a combined index $a \equiv (i,j)$ with $i<j$ and $a=1,\dots,M$ with $M=N(N-1)/2$ . Finally, we have defined the inverse moment of inertia tensor $\mathcal{J}_{ab}=4 J_{ab}/\hbar^2$.

\begin{figure}[t!]
  \centering
  \includegraphics[width=0.95\columnwidth]{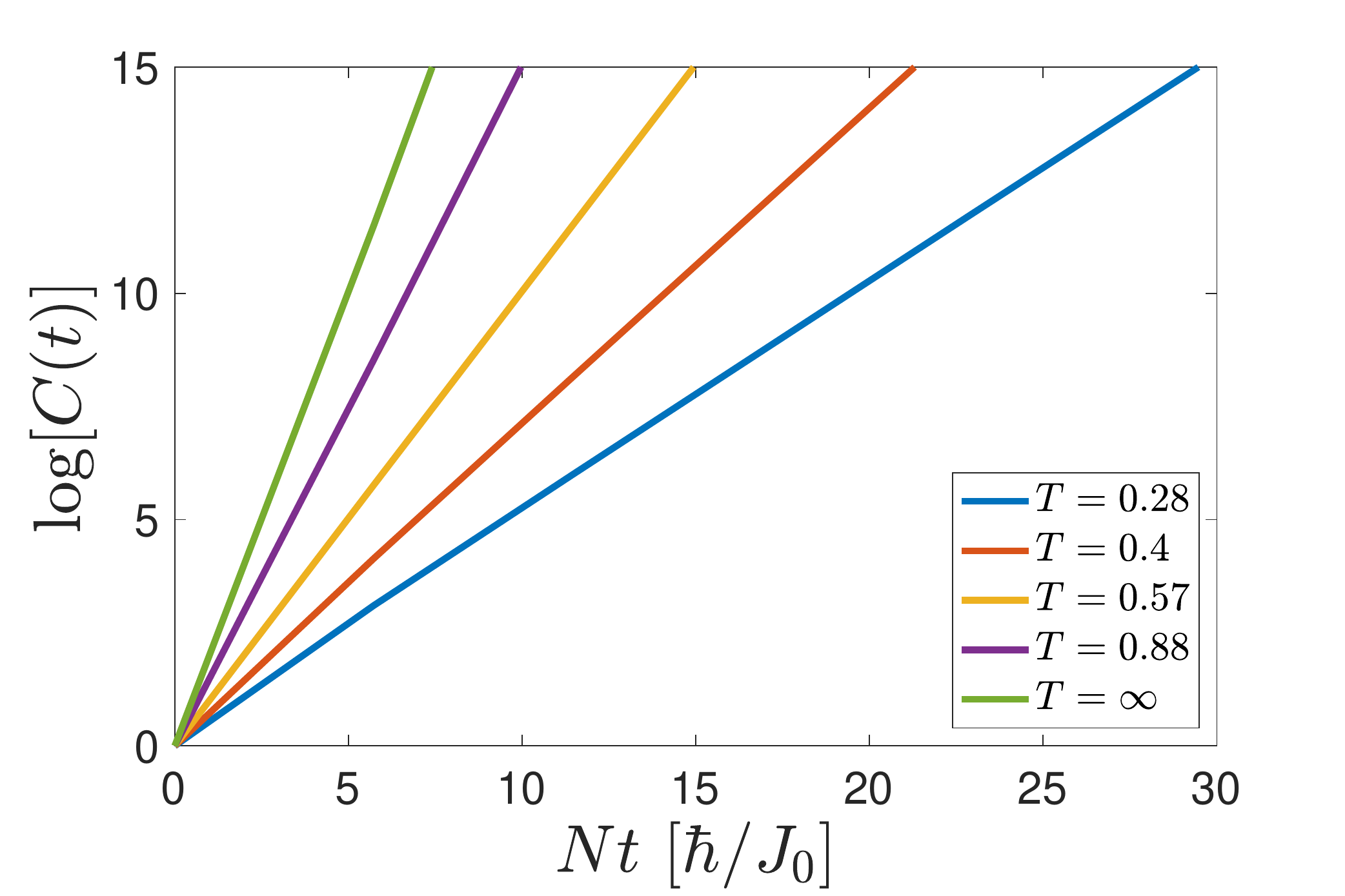}
\caption{Exponential increase of the correlator $C(t)$ at different temperatures. Data were calculated for $N=40$ and $S=1/2$.
\label{Fig1}}
\end{figure}

The above rotor Hamiltonian defines a broader class of quantum models distinguished by the representation of $\mathfrak{so}(N)$ under which the operators $\hat{L}_{a}$ transform. 
The original SYK model corresponds to the $S=1/2$ spinor representation of $\mathfrak{so}(N)$. 
In the following, we will consider the higher-$S$ generalization of this model:
\bea
\hat{H}(S) = \frac12 \sum_{a,b} \mathcal{J}_{ab} \hat{L}_a(S) \hat{L}_b(S) 
\label{SYKHam}
\eea
where the $\hat{L}_a(S)$ transform under the spin-$S$ spinor representation of $\mathfrak{so}(N)$.
This representation is obtained by taking the tensor product of $2S$ copies of the $S=1/2$ representation, and by restricting the Hilbert space to the subspace with the highest value of the quadratic Casimir \cite{simon1980}.
Explicitly, one writes $\hat{L}_{ij}(S) = i \hbar \sum_{\alpha} \gamma_{i,\alpha} \gamma_{j,\alpha}/2$, where $\alpha=1\dots 2S$ is the copy index.

In the $S \to \infty$ limit, the above model leads to the following classical Hamiltonian \cite{simon1980}:
\bea
\mathcal{H}_{cl} = \frac12 \sum_{a,b} \mathcal{J}_{ab} \mathcal{L}_a \mathcal{L}_b 
\label{Hcl}
\eea
where $\mathcal{L}_a$ should be understood as the components $\mathcal{L}_{ij}$ of a $N$ by $N$ anti-symmetric matrix. 
 The Hamiltonian (\ref{Hcl}), along with the Poisson bracket relations $\{\mathcal{L}_{a}, \mathcal{L}_{b} \}_\text{P} = -f_{abc} \mathcal{L}_{c}$ leads 
 to the equations of motion 
\be
\partial_t \mathcal{L}_a =    f_{abc} \mathcal{L}_b  \mathcal{J}_{cd} \mathcal{L}_{d} .
\label{Dynamics}
\ee
We note that similar classical equations of motion appear when studying SYK within the Truncated Wigner Approximation \cite{Davidson2017,PhysRevB.99.134301}.

The dynamics of Eq. 4 can be thought of as the rotation of an $N$-dimensional body with an inertia tensor given by $\mathcal{J}_{ab}^{-1}$ and angular momentum components given by $\mathcal{L}_a$.
There is one important caveat though: there are some strong restrictions on $\mathcal{J}_{ab}^{-1}$ for it to correspond to the inertia tensor of a rigid body, and these restrictions will not generically be satisfied in our model, where $\mathcal{J}_{ab}$ is a random matrix.
This is the reason why $N$-dimensional rigid body rotation is integrable~\cite{1995}, while the dynamics of Eq.~4 will be shown to be chaotic in the next section.
The analogy with rigid body rotation is nevertheless intuitively useful and we will keep using it in the following.

Besides the energy, this problem has another integral of motion of note: the total angular momentum (normalized by the number of components) $P^2 = M^{-1} \sum_a \mathcal{L}_a^2$.
While $P^2$ could take any value for a regular classical system, in our case its value is fixed by the representation of the corresponding quantum problem. In the spin-$S$ spinor case at hand, one finds $P^2 = N S^2 \hbar^2 / 2 M$ ~\footnote{There are actually $N/2$ integrals of motion, given by $C_k=\text{tr}(\mathcal{L}^{2k})$, where $\mathcal{L}$ is understood as an $N$ by $N$ antisymmetric matrix, and where $P^2 \propto C_1$. It was checked that imposing the higher constraints (for $k>1$) to be at their ``quantum value'' does not change the results qualitatively, so we treated them in the canonical ensemble for the sake of simplicity.}.
We decide to scale the couplings such that the bandwidth is $S$-independent: $\overline{\mathcal{J}_{ijkl}^2} = J_0^2 \hbar^{-2} S^{-2}$.
It can be seen from Eq.~\ref{Dynamics} that the actual value of $S$ only has a trivial effect on the classical dynamics: it rescales time by $S$.
All the numerical data are given for $S=1/2$.
 Given these definitions, one finds easily that the energy bandwidth scales like $N^2$, as expected for a classical model with $M \sim N^2$ degrees of freedom.

\begin{figure}[t!]
  \centering
  \includegraphics[width=0.95\columnwidth]{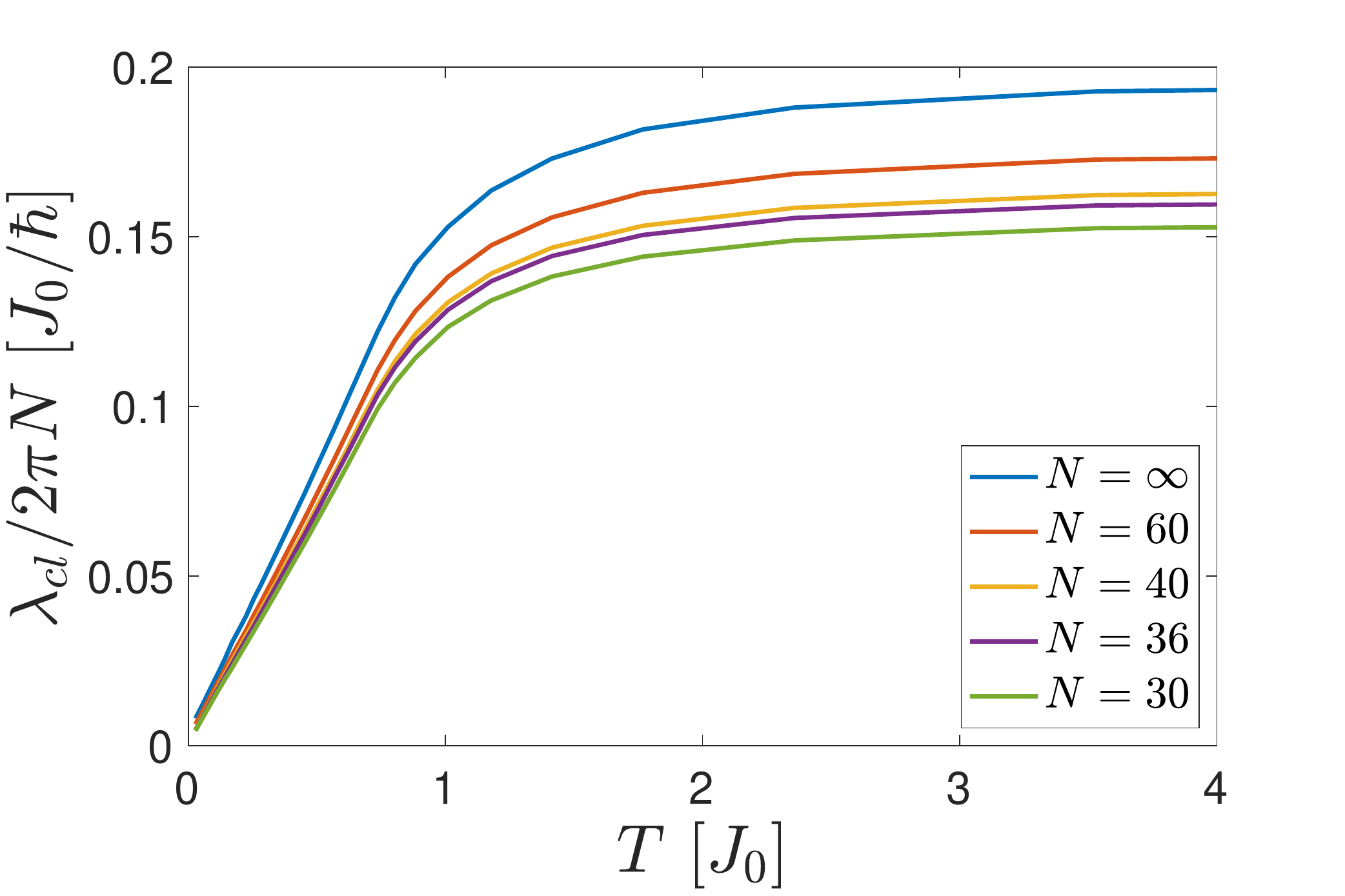}
\caption{Classical Lyapunov exponent versus temperature for $S=1/2$. The $N=\infty$ was obtained by extrapolation, see Fig.~\ref{Fig3}.
\label{Fig2}}
\end{figure}

Finally, we note that the classical Hamiltonian (\ref{Hcl}) with the constraint on $P^2$ maps to the $p$-spin spherical model with $p=2$~ \cite{PhysRevLett.36.1217}.
It has a paramagnetic phase for $T>T_\text{sg}$ and a spin glass phase for $T<T_\text{sg}$, with the spin-glass transition temperature given by $T_\text{sg}/J_0 = 1/\sqrt{2}$.
It is important to emphasize, however, that the Poisson-bracket dynamics we consider for this model is different from previously studied dynamical models involving $p$-spin models \cite{0305-4470-28-15-003,Cugliandolo_2018}.

\section{Classical chaos}
\subsection{Numerical results}

We study numerically the time evolution for the conservative dynamics given in Eq.~\ref{Dynamics}, averaged over an ensemble of initial states $\{ \mathcal{L}_a(0) \}$ drawn from a thermal distribution at temperature $T$ and for the fixed value of $P^2$ given above.

In order to probe chaotic behavior, we compute the following correlation function:
\bea
C(t) = \frac1{M^2}\sum_{a,b}  \overline{\left\langle \left( \frac{\partial \mathcal{L}_a(t)}{\partial \mathcal{L}_b(0)} \right)^2 \right\rangle},
\eea
where the brackets signify average over initial conditions and the overline signifies average over disorder realizations.
This quantity measures the sensitivity to perturbations in the initial condition.
It is the leading term in a semiclassical expansion of the quantum out-of-time-order correlator (OTOC)\cite{larkin1969quasiclassical,COTLER2018318}, and should grow as $C(t) \sim e^{2 \lambda_\text{cl} t}$, where $\lambda_\text{cl}$ is the (largest) classical Lyapunov exponent.
As seen in Fig.~1, $C(t)$ exhibits exponential growth over many decades, and we can reliably extract $\lambda_\text{cl}$ from a linear fit of $\log\left[ C(t) \right]$.

We find that $\lambda_\text{cl}$ grows linearly with $N$ (see Fig. 3).
This may seem surprising for a system with a finite bandwidth for the energy density (defined as the energy per degree of freedom, $E/M$ in this case).
The reason is that each angular momentum component $\mathcal{L}_a$ has a non-zero Poisson bracket with $\sim N$ other components, which is indeed quite unusual: for most systems, the number of degrees of freedom (DOF) having a non-zero Poisson bracket with a given DOF does not scale with the system size.
Combining this scaling with the trivial scaling of Eq.~\ref{Dynamics} with $S$ leads to an overall scaling of $\lambda_\text{cl} \propto N/S$.

Besides this overall scaling with $N/S$, $\lambda_\text{cl}$ shows interesting behavior with temperature (see Fig.~\ref{Fig2}).
While the high-temperature Lyapunov exponent shows saturation with $T$, the low-temperature ($T \ll J_0$) regime is linear in $T$.
The corresponding slope is an interesting quantity since it is independent of the UV scale $J_0$, and it can be compared with the chaos bound of Ref.~\cite{Maldacena2016}:
\bea
2 \lambda_\text{cl} = \eta \frac{N}{2S} \lambda_\text{bound}
\label{MoneyEq}
\eea
with $\eta \simeq 0.31 $ and $\lambda_\text{bound}=2\pi  T/\hbar$. This result was obtained from extrapolating numerical results up to $N=60$ (meaning $M=1770$ classical degrees of freedom), as shown in Fig.~3.

It is quite striking that this purely classical model reproduces a $T$-linear Lyapunov exponent, indicating that such a temperature dependence is not a specific signature of a quantum system, but more of a system with fixed total angular momentum.
The fact that the low-$T$ slope is parametrically different in the classical case than in quantum case raises interesting questions which will be commented on in the discussion.
Note also that this linear in $T$ behavior happens within the thermodynamic spin glass phase ($T<J_0/\sqrt{2}$).

\begin{figure}[t!]
  \centering
  \includegraphics[width=0.95\columnwidth]{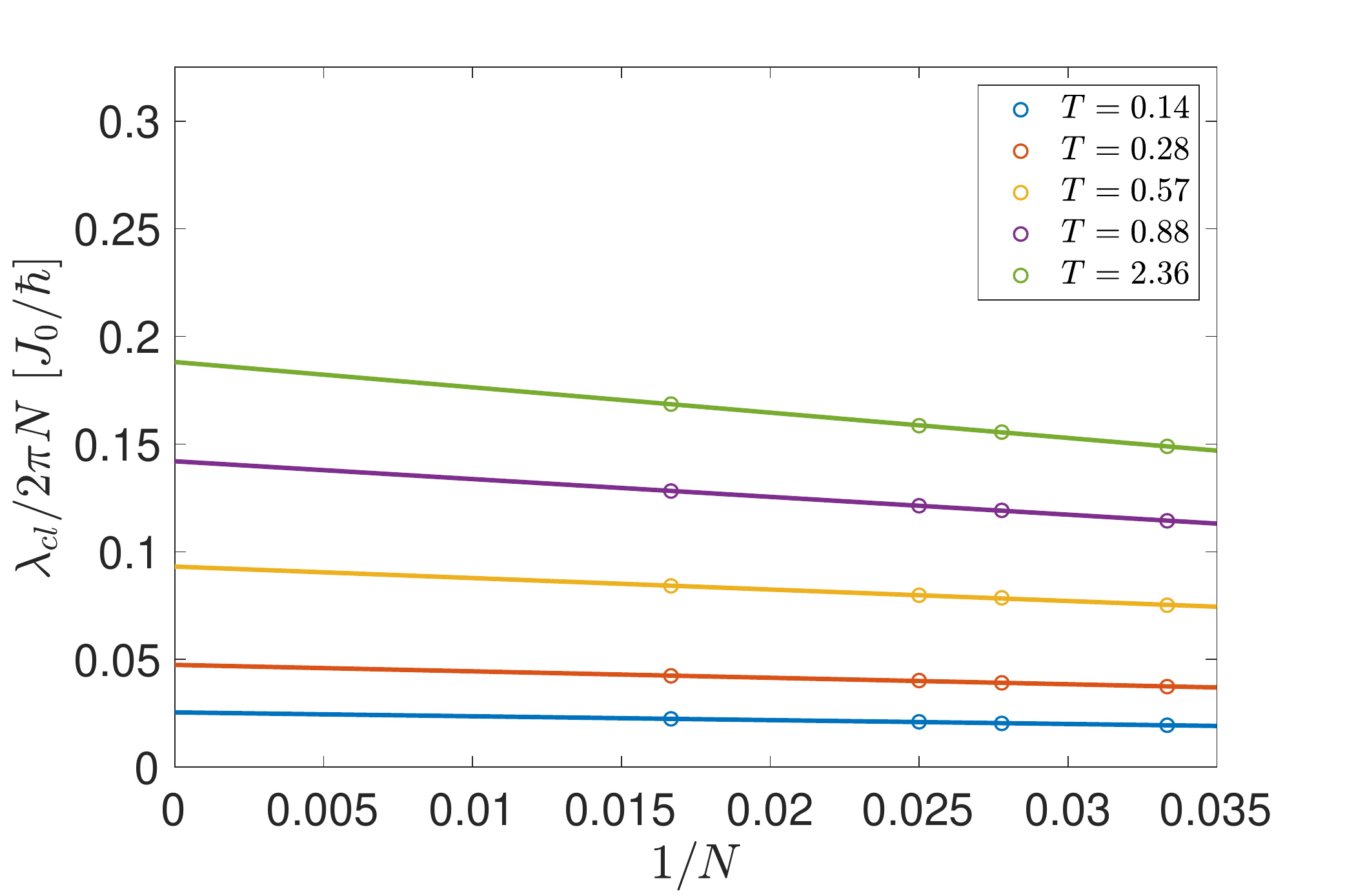}
\caption{$\lambda_\text{cl}/N$ is shown to converge to a finite value in the thermodynamic limit. The circles are numerical data for $N=30,36,40,60$ and $S=1/2$, and the lines are obtained by linear extrapolation. 
\label{Fig3}}
\end{figure}

\subsection{Fixed points and linear stability}
In this section, we identify fixed points of the dynamics, and show how they can be utilized to obtain a spectrum of local Lyapunov exponents \cite{Abarbanel1991}.
It is easy to show that Eq. (\ref{Dynamics}) has $2M$ fixed points given by $\pm \mathcal{L}^{\nu}$, where $\mathcal{L}^{\nu}$ are the (properly normalized) eigenvectors of $\mathcal{J}_{ab}$ with eigenvalues $j^{\nu}$ ($\nu=1,\dots,M$). 
The intuition for these fixed points is clear: they correspond to the body rotating with the maximal angular momentum around one of its principal axes of inertia, defined by the $\mathcal{L}^{\nu}$. 
The energy of the system at these fixed points is given by $E^{\nu}= \frac12  j^{\nu} M P^2$. 
The $j^{\nu}$ obey the semi-circle law and the classical ground (resp. top) state corresponds to the minimal (resp. maximal) value of $j^{\nu}$.
By sweeping over $\nu$, one can therefore analyze the typical behavior over the full range of energy densities.

To perform the linear stability analysis around a given fixed point $\mathcal{L}^{\nu}$, we define the deviation $l$ as $\mathcal{L}(t)=\mathcal{L}^\nu+ l(t)$. We now linearize the equations of motion: 
\bea
\partial_t l_a &= \alpha K_{ab}^{\nu} l_b \\
K_{ab}^{\nu} &=  f_{\nu ac } (  -\mathcal{J} +  j^{\nu}   )_{cb} \equiv f_{\nu ac } \tilde{\mathcal{J}}^{\nu}_{cb}
\eea
where $\alpha=\sqrt{M} P$ (no summation over $\nu$ is implied).
$K^{\nu}_{ab}$ is the $M$ by $M$ stability matrix of the fixed point $\mathcal{L}^{\nu}$ and $j^{\nu}$ is the eigenvalue of $\mathcal{J}$ corresponding to the eigenvector $\mathcal{L}^{\nu}$.

One should now analyze the spectrum of $K^{\nu}$: imaginary eigenvalues correspond to oscillatory behavior while real positive (resp. negative) eigenvalues correspond to unstable (resp. stable) directions, and can be interpreted as local Lyapunov exponents.
It is instructive to first consider the case of dissipative (steepest-descent) dynamics \cite{0305-4470-28-15-003}, where $\mathcal{L}(t)$ follows the gradient of the energy, subject to the constraint of keeping $P^2$ fixed.
This dynamics has the same fixed points as the Poisson bracket dynamics (i.e. $\pm \mathcal{L}^{\nu}$), but the stability matrix is instead given by $\tilde{\mathcal{J}}^\nu$.
Using the fact that $\tilde{\mathcal{J}}^\nu$ is a real symmetric matrix, 
one finds easily that there are only stable and unstable directions (no oscillatory behavior), and that the number of unstable directions goes from 0 in the ground state to $M$ in the top state.
This means that the ground state is a global attractive fixed point, the top state a global repulsive fixed point, and the fixed points in the middle are saddles with an index that interpolates between the two cases.
This type of dynamics, associated with random thermal noise, was solved in Ref.~\cite{0305-4470-28-15-003} and was shown to have glassy behavior at low $T$.

\begin{figure}
  \centering
  \includegraphics[width=0.95\columnwidth]{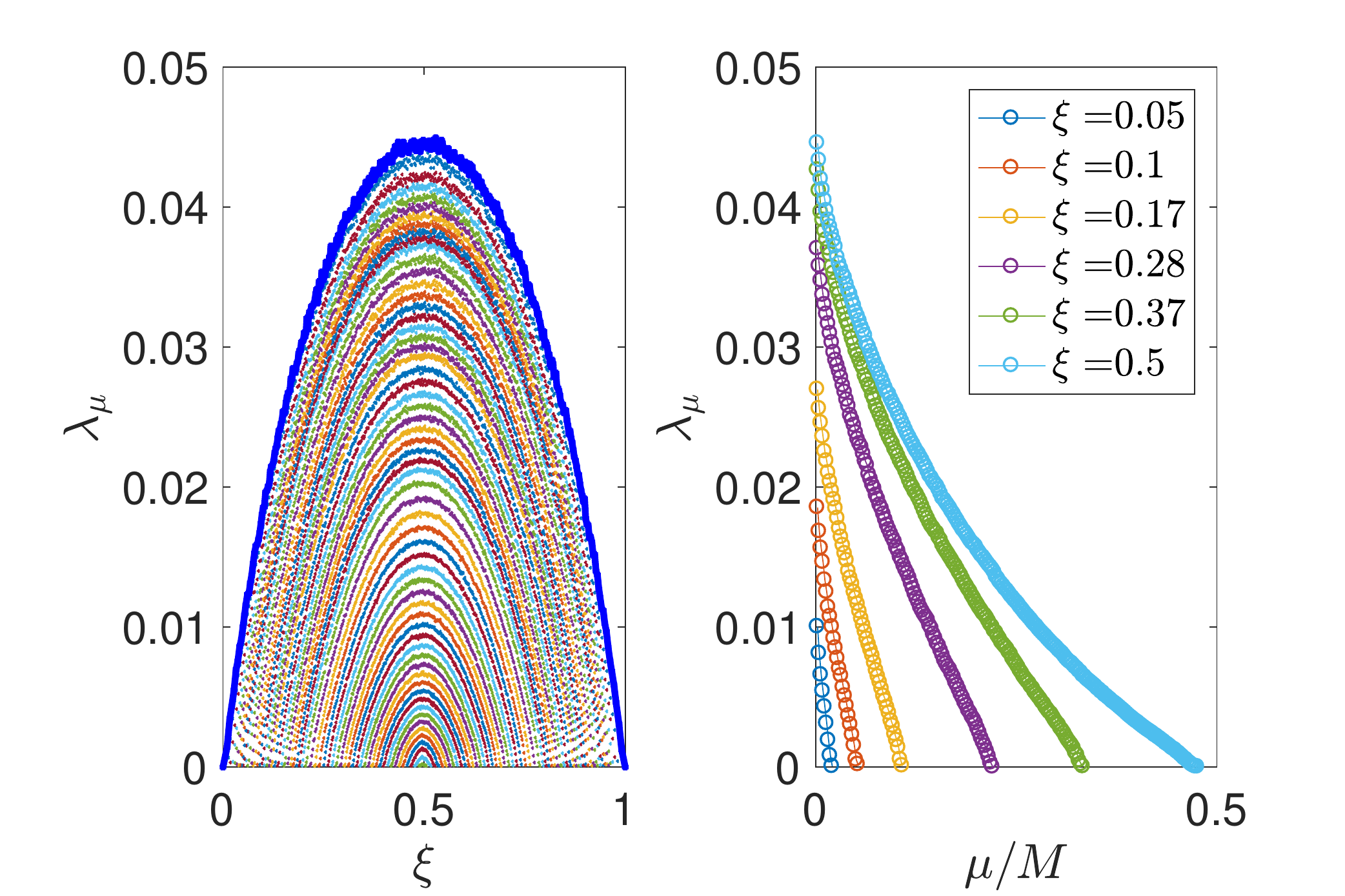}
\caption{Left panel: Spectrum of local Lyapunov exponents versus energy density $\xi=(E-E_\text{GS})/(E_\text{TS} - E_\text{GS})$, with $E_{GS}$ the ground state energy and $E_{TS}$ the top state energy. Averaged over 50 disorder realizations for $N=2S=30$. Right panel: Same spectrum but plotted against spectrum index $\mu$ for fixed values of $\xi$.
\label{LyapunovSpectrum}}
\end{figure}

Instead, for the Poisson bracket dynamics, $\tilde{\mathcal{J}}^\nu$ is multiplied by an antisymmetric matrix $f$ in order to obtain the stability matrix. 
This leads to a dramatic change in the dynamics which can be understood as follows.
 Let $K=AB$ with $A$ an antisymmetric matrix and $B$ a symmetric matrix. 
 If $B$  is positive definite or negative definite then the spectrum of $K$ is purely imaginary. 
 This means the stability matrix around the ground and top states has a purely imaginary spectrum for Poisson bracket dynamics.
 The motion around the ground state (and the top state) is thus purely oscillatory, in contrast to the dissipative dynamics case for which the ground (resp. top) state only had stable (resp. unstable) directions.
As one considers fixed points away from the ground (or top) state, $\tilde{\mathcal{J}}^\nu$ becomes less and less definite, leading to more and more $K$ eigenvalues with non-zero real part.
These real eigenvalues, when positive, can be interpreted as local Lyapunov exponents, and their number and size increases as one moves away from the ground or top state, as shown in Fig.~\ref{LyapunovSpectrum}.

\section{Discussion}

In conclusion, we have introduced a generalization of the SYK model by introducing $2S$ species of fermions.
We have studied the infinite-$S$ limit of this model, a classical model whose dynamics can be understood as the rotation of an $N$-dimensional body.
The temperature dependence of the Lyapunov exponent was studied numerically, and a $T$-linear dependence was found at low $T$, like in the quantum case.
The corresponding slope was found to scale like $N/S$.
The thermodynamics maps to the $p=2$-spin model, whose spin glass phase at low-$T$ seems to have little effect on chaos.
Finally, we have identified an extensive number of fixed points, which enabled us to compute the spectrum of local Lyapunov exponents in their vicinity.

It is natural to ask how quantum fluctuations affect chaos when going from the classical model of Eq.~\ref{Hcl} to the quantum model of Eq.~\ref{SYKHam}.
For many toy models of single-particle chaos, like quantum billiards, this question is typically answered geometrically: the exponential sensitivity to initial conditions survives as long as the de Broglie wavelength is much smaller than the relevant ``chaos length scale'', like the radius of curvature of the billiard.
In fact, an interpretation of the quantum bound on chaos \cite{Maldacena2016} was given in those terms in Ref.~\cite{2016arXiv161201278K}.
However, it is not a priori clear how to transpose this analysis to a system of interacting quantum fermions, which do not generically have such a geometrical picture.

Interestingly, the classical dynamics obtained here can be reformulated geometrically: as shown by Arnold\cite{Arnold}, it corresponds to a particle following geodesics on a $SO(N)$ manifold equipped with a metric related to the inertia tensor of the body.
Chaos can then be understood as arising from diverging geodesics due to the non-zero curvature of the manifold\cite{CASETTI2000237}.
We surmise that this geometrical formulation of many-body classical chaos could provide a starting point for an analysis of the classical to quantum crossover, whereby the de Broglie wavelength is compared with the different radii of curvature.
Furthermore, the randomness of the couplings in SYK would lead to a random metric, whose properties could be studied statistically.

It is worth noting that, even at the purely classical level, the model we have considered bears surprising similarities to the universal physics of quantum fast scramblers, such as the SYK model. First, it show a linear dependence of the Lyapunov exponent. Second, it possesses a time reparametrization symmetry similar to that found in the quantum SYK model \cite{PhysRevD.94.106002}, since the problem of finding geodesics is explicitly time-independent \cite{2016arXiv161201278K}.

The question remains of whether these are generic properties of the classical limit of bound-saturating quantum systems.
In this context, a straightforward extension of our work would be to study the classical limit of other fast scramblers, like the SYK model with $q$-body interactions, with $q>4$, and the disorder-free tensor models of Refs.\cite{2016arXiv161009758W,PhysRevD.95.046004}. 
Finally, the possible interplay between the chaotic dynamics reported here and the glass phase of the $p$-spin models should also be investigated.

\begin{acknowledgements}
We thank Alexander Altland, Sumilan Banerjee, Xiangyu Cao, Victor Galitski, Sriram Ganeshan, Antoine Georges, Yingfei Gu, Alex Kamenev, Anatoli Polkovnikov and Ari Turner for insightful discussions. 
We acknowledge support from the Emergent Phenomena in Quantum Systems initiative of the Gordon and Betty Moore Foundation (T.S.), ERC Synergy grant UQUAM, DOE Office of Science, Office of High Energy Physics grant DE-SC0019380, and the Mike Gyorgy Chair in Physics at UC Berkeley (E. A.).
 The numerical computations were carried out on the Lawrencium cluster resource provided by the IT Division at the Lawrence Berkeley National Laboratory under the DOE contract 15DE-AC02-05CH11231.
\end{acknowledgements}

\bibliography{ClassicalLargeN_paper}

\end{document}